# Alternative Authentication in the Wild

Joseph Maguire & Karen Renaud
School of Computing Science
University of Glasgow
Glasgow, Scotland, UK
Email: {joseph.maguire,karen.renaud}@glasgow.ac.uk

*Abstract*—Alphanumeric authentication routinely fails to regulate access to resources with the required stringency, primarily due to usability issues. Initial deployment did not reveal the problems of passwords; deep and profound flaws only emerged once passwords were deployed in the wild. The need for a replacement is widely acknowledged yet despite over a decade of research into knowledge-based alternatives, few, if any, have been adopted by industry.

Alternatives are unconvincing for three primary reasons. The *first* is that alternatives are rarely investigated beyond the initial proposal, with only the results from a constrained lab test provided to convince adopters of their viability. The *second* is that alternatives are seldom tested realistically where the authenticator mediates access to something of value. The *third* is that the testing rarely varies the device or context beyond that initially targeted. In the modern world different devices are used across a variety of contexts. What works well in one context may easily fail in another.

Consequently, the contribution of this paper is an "in the wild" evaluation of an alternative authentication mechanism that had demonstrated promise in its lab evaluation. In the field test the mechanism was deployed to actual users to regulate access to an application in a context beyond that initially proposed. The performance of the mechanism is reported and discussed. We conclude by reflecting on the value of field evaluations of alternative authentication mechanisms.

## I. INTRODUCTION

Passwords are not only powerful in theory, but are the most popular form of authentication in practice [1]. The dominance of passwords is unsurprising, given the numerous advantages of the approach [2]. The concept is inexpensive to implement, trivial to learn and works across countless devices [3]. Nevertheless, despite such strengths, passwords are plagued with usability problems that ensure they are increasingly not fit-for-purpose [4], [5], [6]. This reality is particularly perplexing given extensive research into knowledge-based alternatives [7].

Dunphy *et al.* [8] argue that the scarcity of deployment evidence and real-world use are prominent explanations for poor adoption of alternatives by industry. Similarly, the emerging trend in alternative authentication research is to propose novel mechanisms but not to conduct extensive exploration and field investigations of actual use [9]. Furthermore, Beautement and Sasse suggest that not only is real-world performance rarely investigated, but data reported from controlled laboratory studies is unlikely to be representative of actual performance [10]. Moreover, investigations rarely consider actual authentication context, target specific task scenarios or even reflect emerging consumer and industry trends [11]. The reality is that individuals may perform the same task, but not necessarily on the same device. An individual may authenticate on a traditional desktop computer during a task on one occasion, but on another occasion they perform the same task and authentication step on a different device, such as a smartphone.

Therefore, any alternative authentication scheme also encounters the challenge of scaling between devices. Passwords have demonstrated an ability to scale between traditional computers, such as desktops and laptops, and more mobile devices, such as smartphones and tablets. This suggests that any alternative authentication solution seeking adoption would either need to demonstrate convincing performance in switching between devices or, at least, outline how solutions would perform when authentication occurs on a non-target device.

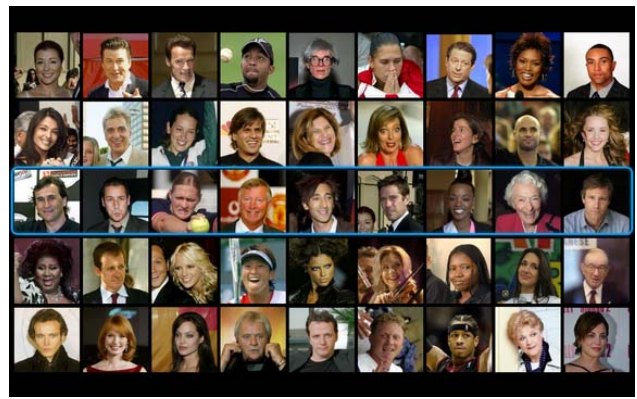

Fig. 1. The previously proposed alternative authentication mechanism, Tetrad. The mechanism was initially designed and targeted for display in a shared-space [12], e.g. television in a living room.

Consequently, the aim of this paper is to disseminate experience of a field investigation of a previously proposed alternative mechanism, in a context outside that originally proposed [12]. This means that authentication was tested *realistically*. It mediated access to something the user cared about: "in the wild".

The **contributions** of this paper are:

- we report on a field investigation of a previously proposed alternative authentication mechanism that performed well it its initial lab-based evaluation;

- we highlight aspects related to the design and methodological considerations that need to be contemplated in carrying out a field investigation of an alternative authentication mechanism;

- we discuss the lessons we learned and reflect on the value of probationing an alternative authentication mechanism *in the wild*.

The expectation is that researchers can build on these lessons to guide their future research and subsequently improve knowledge transfer and adoption of promising alternatives by industry.

## II. BACKGROUND LITERATURE

### A. Alternative Authentication

Researchers have proposed many alternative authentication mechanisms, as reviewed by Biddle *et al.* [1]. Blonder made the first notable proposal [13], it required individuals to remember positions within an image. PassPoints, proposed by [14], is built along the same lines as Blonder's patent. Passpoints was evaluated by 40 evaluators but not as part of a realistic workflow.

Draw-a-secret or DAS, proposed by [15], is a recall-based graphical authentication mechanism requiring an individual to draw an authentication secret to access an application.

Another variation of alternative mechanisms is those that require users to identify "their" secret images from one or more challenge sets. PassFaces is one of the few commercial mechanisms of this kind. The authentication approach assigns an individual a collection of faces as their authentication secret. One using abstract images was proposed by [16], called Déjà Vu. Others have used handwritten numerals [17], Mikons [18] and cultural images [19].

Graphical authentication mechanisms are highly susceptible to onlooker observation. The majority of designers implicitly assume that all interaction with their mechanisms occurs in solitude. The beauty of alphanumeric authentication is that it relies on specific hardware that can, for the most part, be wielded privately. However, this does not mean that alphanumeric authentication, as a process, is somehow superior in terms of resisting observation. A number of mechanisms have attempted to resist observation [20], [21], [22], [12]. We will be trialling one of these in our field test.

### B. Field Testing

No one would dream of releasing a product without carrying out tests with end users. The testing of user interfaces is often carried out in a lab-based situation using well established techniques from the field of psychology. Yet Thomas and Kellogg [23] argue that such studies are severely limited due to omissions of real-life factors and additions of other factors due to the tightly controlled evaluation environment. They explain that the evaluation is not ecologically sound.

A field test is essentially a half-way house between a strictly controlled evaluation and a launch to widespread use in a variety of contexts, by a diverse user base. One tests the mechanism outside a controlled environment "in the wild", relinquishing control to see how it will perform once it emerges from a protected environment.

It has been argued that field testing, is essential, that one cannot stop at a successful lab test and trumpet successful results if performance is promising [2], [10]. The need for field testing is a strong requirement, and does not necessarily enjoy widespread agreement. Kjeldskov *et al.* [24], for example, argue that field studies of mobile interfaces, is not worth the time, expense or hassle. Similarly, Kaikkonen *et al.* [25] found that studies delivered little benefit based on the time and expense of field trial.

Yet, Nielson *et al.* [26] declare field testing worthwhile. They found that issues such as cognitive load and interaction style were only discovered during the field study of a mobile interface. Furthermore, Rogers *et al.* [27] argue that lessons learned in field tests are essential in refining applications and improving the user experience. Bannon [28] argues for field investigations as follows: *"What is needed are both 'quick and dirty' methods that can give rapid feedback ... as well as more extensive field evaluations ... before, during and after the adoption of new technologies"* (p.233).

### C. Field Testing of Alternative Authentication

The unfortunate reality is that despite a decade and a half of research into graphical authentication mechanisms industry has not embraced them. There may be many reasons but one of the most obvious is that industry is not convinced that something hatched in a hot house will be able to perform its function when subjected to the strains of widespread use by a heterogeneous user base.

Most well-known alternative authentication mechanisms have undergone only small-scale or lab-based evaluations, eg. [29], [16], [30]. Some notable exceptions are worth mentioning. Brostoff & Sasse performed a field-study of Passfaces [11], one of few commercially available mechanisms [31]. In terms of context, participants were students, using the mechanism as part of a Web application for coursework. They measured how long authentication took and found that it took a lot longer than password authentication and this caused their participants to use the system less often. Davis *et al.* [32] uncovered some problems with respect to predictability of choice of faces, mirroring the predictability issues of passwords themselves.

Chiasson *et al.* [33] also performed a field-study. The alternative that was trialled was PassPoints, a cued-recall approach devised by Wiedenbeck *et al.* [34], [14]. The study also used Computing Science undergraduates to evaluate the mechanism. The initial work unearthed efficiency concerns, with times required to enter the secret being much longer than password entry times. Chiasson *et al.* reported that authentication improved with practice and that students were incredibly accurate: 78% of attempts were within 4 pixels or 1.5mm. Many in industry would nou doubt consider a 78% success rate to be unacceptable. In the study by Shay *et al.* [35] more than half of their participants dropped out during their evaluation. No real-life service can risk annoying their customers this much.

## III. AUTHENTICATION IN THE WILD

The aim of assessing an experimental authentication mechanism in the wild is to gather evidence of performance and acceptance by individuals. Such an assessment acknowledges the reality that while an experimental security solution may

33

perform well in a controlled setting, it may perform dramatically differently in the wild [36]. Performance in a particular environment is not necessarily indicative of performance in another [37]. The motivation for assessment in the wild is that any evidence gathered during "in the wild" deployment will be indicative of realistic performance. Nevertheless, various aspects of evaluation in the wild need to be given due consideration in order to ensure that the evaluation is not only fruitful but representative of real-world performance.

The first aspect to consider is the *ecological validity* of the evaluation. An authentication mechanism is typically deployed in the wild, as a preliminary to the actual task of interest. The task is of primary importance; authentication is a secondary goal: another step among many [10]. Consequently, authentication in the wild can be only realistically be considered as part and parcel of a package; broadly comprising an *authentication mechanism* and associated *application*.

The authentication mechanism must sit comfortably within such a package. The authentication mechanism and application should represent a balance in terms of security needs [38]. It seems senseless, for example, to embed military-grade authentication within a consumer-focused task. An unrealistic coupling is likely to lead consumers to either abandon the application or resort to undesirable coping mechanisms. Matching the strength of the mechanism to the value of the protected asset is an important consideration not only during deployment, but in a strong ecologically valid evaluation as well.

The *persuasiveness* of the primary task is another important aspect to consider in terms of field evaluation. The reality is that many assessments of authentication mechanisms rely on perverse incentives or unrealistic incentives, e.g. course credit. These incentives could potentially influence frequency of use and mask the acceptance of an authentication mechanism. If we avoid incentives we have to ensure that the primary task is persuasive enough to motivate consumers to return to the application by offering actual benefits [39]. The barrier to their receiving such benefits is the experimental authentication mechanism. A costly complex solution would arguably lead to a decrease in authentication attempts over time.

An authentication mechanism must have a *clear purpose*, and this purpose should be transparent to the user. This is an important aspect to consider as users should be empowered with a clear understanding of the *consequences* of authentication [40]. The presentation of the authentication mechanism should be consistent and transparent to ensure users are aware of the outcome of authenticating. The concern is that inconsistent and thoughtless presentation of the authentication mechanism may confuse users, influencing their decision to authenticate, or not. Moreover, thoughtless and repeated authentication could be exploited by attackers since users may become desensitised and simply authenticate without questioning the consequences of authenticating.

In summary, there are at least three important aspects to address when evaluating an experimental authentication mechanism in wild, they are:

- *Ecologically-valid application*
  An authentication mechanism is coupled with a realistic primary task [36], e.g. military-grade authentication solution is evaluated as part of a military-grade task.

- *Persuasive primary task*
  The application should offer benefits and avoid perverse incentives that could influence behaviour [39], i.e. course credit and monetary rewards are significant benefits, potentially worth the cost of complex authentication.

- *Clear consequence of authentication*
  The authentication mechanism should be thoughtfully presented to ensure the user has a clear perception of the outcome of authentication [40].

Nevertheless, while the aforementioned aspects are important to the exploration of experimental authentication mechanisms in the wild, there are also other pertinent issues that need to be considered. Researchers have a responsibility for individuals involved in any investigation. Consequently, application use must not harm [41], as it is unacceptable for any investigation to impact on users in a negative way. In the context of evaluating an authentication mechanism, an advisable direction is to use a 'low risk' application, that collects minimal information for the purposes of evaluation [42].

There is also the responsibility of managing the transition and/or sustainability of the application [43]. If the application does deliver tangible benefits and participants come to depend upon it, researchers should support a careful transition. In some senses this is important in order to sustain future research and investigation but there are also ethical implications. If, for example, students come to depend on an application during their studies, withdrawing it or failing to patch it could negatively impact on their performance i.e. inadvertently cause harm.

Consequently, in application deployment the follow areas need to be given due consideration as well:

- *Ethical responsibility for participants*
  The desire to evaluate an alternative authentication mechanism in a realistic setting must not override the safety or well-being of the user [41].

- *Ecological responsibility for participants*
  The application should exist beyond the evaluation period, as participants could come to depend on the application or the ecology supporting it [43].

These ethical and ecological aspects along with the aforementioned aspects pertinent to authentication research were considered and influence the design of our investigation.

## IV. METHODOLOGY

### A. Participants

The participants were enrolled undergraduate students at the School of Psychology, University of Glasgow. The demographic was diverse but only students owning an Apple iPhone and valid iTunes Store account could download and execute the application.

*B. Apparatus & Measurements*

A bespoke iOS application was developed and designed for participants to download, listen and annotate lecture recordings. An arguably persuaisive primary task for student enrolled in a course. The use of the authentication was clear and consistent: access to the application was meditated with Tetrad, participants authenticated to access it. The application could be downloaded, installed and executed on any device with iOS 4.0 or later. The application was not designed for Apple iPad but could execute on the device, if operating system requirements were met. The package was considered an ecologically valid application of Tetrad, as the mechanism was envisioned as part of a digital content purchasing process.

A registration code was requested from participants when the application was first launched. The application was not usable without a registration code, issued in conjunction with the School of Psychology. Registration codes could only be used once and if an individual had to re-register, they had to request an additional registration code. The registration process did not collect any personal information from participants. The registration code was purely for the creation of a Tetrad authentication secret.

The alternative authentication mechanism was bootstrapped with an image-set of staff photographs provided by the School of Psychology. An image-set of Hollywood celebrities was initially favoured due to their perceived familiarity with participants, but complicated due to legal and technical concerns. Consequently, the image-set of staff photographs was considered a close approximation as staff are 'community celebrities', i.e. familiar among students. Furthermore, legal and technical issues could be addressed as permission could be sought directly from staff and the School itself.

Metrics pertaining to performance were recorded and collected for registration and authentication processes. The key metrics include the time spent on registration and authentication as well as the success and alignment of each authentication attempt.

These metrics were logged to assess the alternative authentication mechanism. The application was considered low-risk, as minimal information was collected and only lecture content was primarily at risk. In order to cater to enrolled students who did not own an iOS device, had no interest in the application *or* had become frustrated with it, a web application that offered similar functionality was deployed. The web application was managed independently and relied on alphanumeric authentication.

Finally, the School of Psychology was informed the alternative authentication mechanism may fail or exhibit errors. The client was content to proceed as only the lecture recordings were at risk, not participants. The proviso was if the application exhibited such errors, it was to be removed from distribution until such as errors were addressed. Lastly, a sustainability agreement was reached that the School of Psychology was allowed use of the application beyond the investigation.

*C. Procedure*

The School of Psychology informed all enrolled students, by email, that an application for iOS devices would be made available during the first term. The communication requested all students to complete a survey that probed the popularity of Internet-connected devices. There was no requirement to participate in the survey, although only those students who had completed the survey and indicated they owned an iOS device would be automatically contacted with a registration code. The application was distribution via the Apple App Store (iOS) and there was no requirement or expectation to use the application. Individuals preferring not to use the application or without an iOS device were advised of the web application. Lastly, participants were advised they could consult a website for documentation, email for technical support and were encouraged to provide feedback via email or iTunes Store review mechanism.

V. RESULTS

*A. Registration*

There were total of 63 registration records, 28 active and 35 inactive, generated by 45 distinct devices. The 63 registration records can be decomposed into three separate groups, in terms of how many times a device was used to register. The device identifier is recorded during the registration stage. The first group is for devices that were used once and comprises 29 registrations records, with 16 active. The second group is for devices that were used twice and comprises 28 registration records, with 14 active. The third group is for devices that were used three times and comprises of 6 registration records, with 1 active. This means 46% of registration records come from devices used to register once. There is also not a diverse selection in images, with users making similar image choices, see Table I.

The average time to complete the registration stage was 312s, with a minimum time of 25.17s and maximum time of 5041.94s. If we consider only active records, average registration time increases to 360.73s while the minimum and maximum time remain the same. If we consider only non-active records, average time decreases to 273.01s with a minimum time of 41.78s and maximum time of 2738.42s.

*B. Authentication*

There are total of 387 attempts, 226 were successful, 161 unsuccessful. The 226 successful attempts can be decomposed into 180 horizontal attempts and 46 vertical image alignments.

The average time for all attempts was 1094.75s, with an average time of 671.68s for all successful attempts and 1688.64s for unsuccessful attempts. The average time for all horizontal attempts was 678.09s and for all vertical attempts was 646.58s.

| Title | P1 | P2 | P3 | P4 | Count |
|---|---|---|---|---|---|
| Senior University Teacher | 12 | 6 | 6 | 11 | 35 |
| Teaching Assistant | 10 | 10 | 4 | 8 | 32 |
| Professor | 9 | 4 | 7 | 8 | 28 |
| Reader | 7 | 10 | 7 | 4 | 28 |
| University Teacher | 5 | 6 | 6 | 5 | 22 |

TABLE I. MOST POPULAR SELECTIONS

The minimum time of 1.43s seems small. If we consider all attempts less than 25s, the average time of an attempt is 13.40s with a minimum time of 1.43s and maximum time of 24.50s.

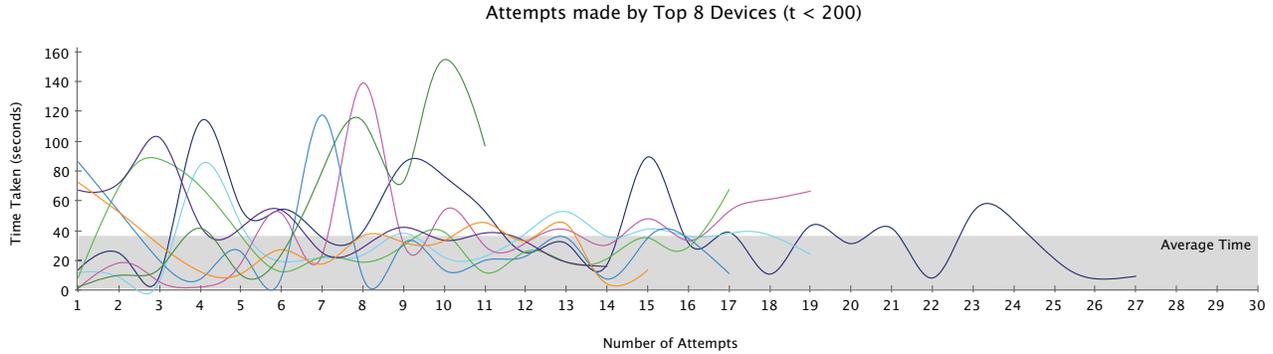

Fig. 2. Attempts made by Top 8 Devices

The average time for a successful attempt is 16.84s with a minimum time of 4.81s and maximum time of 24.21s, leaving 47 records. The average time for an unsuccessful attempt is 11.86s with a minimum time of 1.43s and maximum time of 24.50s, leaving 101 records.

There are 52 attempts taking less than 10 seconds, with an average time of 6.81s. There are 7 successful attempts, all horizontal, with an average time of 8.02s. There are 45 unsuccessful attempts with average time of 6.62s, with a minimum of 1.43s and maximum of 9.86s. This means that 68.2% of attempts took less than 25 seconds and 86.5% of attempts less than 10 seconds are unsuccessful.

## VI. DISCUSSION

The evaluation of a graphical authentication mechanism with non-technical students is relatively unusual as most mechanisms are deployed to technical students, enrolled in science and engineering courses [11], [33], [44]. Furthermore, targeting the mechanism at smartphone users was fertile ground as the application was installed on a respectable number of devices, 45 out of a possible 90.

It seemed to make sense to trial Tetrad's new Smartphone implementation, since it had performed well in lab-based evaluations. Based on its probation we have to conclude that Tetrad is not a viable mechanism. A number of problems emerged during the field trial.

### A. Effectiveness

The application was downloaded, installed and used by many students, much like any other mobile application. Usage logs revealed that the community mainly used the application during pre-exam revision. On reflection, the application was mediocre for the community it served. This is a valuable lesson as proper consideration of community activity would have not only made for a better application, benefiting students, but would have generated more usage information to support meaningful analysis and a more realistic probation of Tetrad.

The mechanism did not demonstrate strength either. Users created predictable passwords by picking the same images, arguably due to the familiarity of some of the images used within the mechanism. This is addressable, if a suitable image set can be identified, but this problem is experienced by most recognition-based graphical authentication mechanisms, and is not specific to Tetrad.

The decision to use staff images was to: (a) harness the desirable aspects of faces by using *community celebrities* while avoiding undesirable aspects such as those reported by Davis *et al.* [32]; and (b) improve communication of the core intent of the mechanism; students would know when this mechanism was presented that they would be authenticating for a task associated with Psychology.

Unfortunately, the use of staff images did not curb the human tendency to create predictable passwords. 58% of image selections were represented by just five images, or rather, five members of staff. These members of staff were all central to teaching within the department, specifically in the early years. Therefore, it seems users were not making assessments necessarily based on attraction but on familiarity.

This finding serves only to confirm the biggest issue of recognition-based graphical authentication approaches: 'bootstrapping' a mechanism with images is challenging [8]. This concern is not just restricted to image type but to image distribution and delivery, as well. The high-resolution displays now present on most '*post-PC*' devices demand high-resolution images, resulting in larger file sizes. These images can be distributed inside the application but this restricts authentication options and increases the size of the application.

Knowing the connection type and data transfer speeds would be valuable in designing future mechanisms. In retrospect, the 'semi-wild' environment used in our probation would have been perfect for extracting such information. Therefore, we advise researchers assessing a mechanism in a similar fashion to log such information.

### B. Satisfaction

The criticism directed at the authentication mechanism took two forms. The first was that some students felt the authentication mechanism was unnecessary. They felt that an application that delivered lecture recordings to an individuals mobile device should be free from the shackles of authentication. Consequently, some users of the application felt that the authentication mechanism was an unrealistic addition. However, such an argument can be contrasted with the requirements laid out by the client. The School of Psychology was adamant that the application access be mediated via an authentication mechanism, during and after the evaluation period. This confirms the difficulty of fixing on a realistic use of an authentication mechanism. The second form of criticism was that the authentication mechanism was unusable. There were some anecdotal comments that suggested the graphical authentication was not well received.

Figure 3 shows a review from the iTunes Store that a student posted after they had downloaded and used the application. It encapsulates both forms of criticism. The comments are visceral and emphasise that the authentication mechanism essentially drove the reviewer away. The authentication approach was novel to the user and this novelty has many connotations, not all positive. The review provides some insight into these negative connotations. The reviewer is arguably distrusting of the authentication mechanism and not convinced that they entered the authentication secret incorrectly or whether the authentication mechanism itself processed it incorrectly. Alternatively, the comment does make a valid user interface remark that the user is unable to make corrections as they are not sure what they have entered. The final remark is that the reviewer is resolved not use the application until the authentication mechanism is removed. The remark makes the point that he/she was not going to waste time dealing with an experimental authentication mechanism, as something untested would not be used to protect something valuable.

The user base was intelligent and arguably did not appreciate being used as guinea pigs to evaluate an authentication mechanism to access valuable learning resources. However, in fairness, students were not forced to use the application and could use a password-protected portal to access lecture recordings. The authentication mechanism clearly alienated some students. The reality is that several users (23%) abandoned the application after authentication and did not return. Therefore, it many respects the alternative authentication mechanism was a failure from the outset. It frustrated too many, and drove too many users away.

*C. Efficiency*

Users spent an excessive amount of time entering secrets, on average 1094.75s. This can be improved by removing outliers, resulting in 36.66s, or 45.02s, if only considering successful attempts. Some shorter successful attempts were made, with the fastest being 4.85s. Unfortunately, users did not seem to improve with successive use. Figure 2 illustrates this, initial oscillation is likely from users becoming familiar with the mechanism. However, even after many attempts oscillation still persists, although it is tempered. This is likely the result of the mechanism being a *searchmetric* [45] and requiring individuals to reposition images once they have recognised them. The reality is that this combination is not best suited to a mobile device, 45.02s is not unforgivably long for an authentication mechanism but is clearly far more time-consuming than password entry.

*D. Interface Inconsistency*

Nearly two thirds of unsuccessful attempts took less than 25 seconds, with most of them taking less than 10s. This unacceptably low success rate suggested a major flaw in the implementation. 41.6% of attempts with the mechanism were unsuccessful and 226 successful.

Upon examination we realised that the problems started at registration. Registration is when a user is first exposed to an authentication mechanism, making it more than just a sign-up phase, it constitutes a learning phase. The interaction a user has with the mechanism during registration is carried with them into authentication. That we failed to understand or appreciate this nuance is our biggest error. Passwords display particular consistency here: type to register, type to authenticate. Our mechanism did not display this consistency.

The numerous, short, failed authentication attempts represented interaction errors rather than memorability errors. During registration, users were required to choose four "secret" images. They did this by double-tapping on the images they wanted. At authentication, they were required by the mechanism to reposition images by swiping left, right, up or down. When images were lined up satisfactorily, they did a double-tap to submit their attempt. There were instructions to this effect.

In our minds the intention was clear: double-tap to submit a choice. During registration, submit your secret image choice, and at authentication submit your image alignment. Unfortunately, what was communicated was: when you see an image, double tap it. We *told* them, in the instructions, about swiping, but their prior experience was more compelling than the displayed instructions and overrode them.

Unfortunately, what happened was that the registration experience habituated them into double tapping on their image of choice. When authenticating, it was natural instinct to see their image and double tap it. This submitted an authentication attempt, which obviously failed since they had not yet aligned their images. It seems that they were unaware of their error, and continued to try to authenticate in the same way, with growing frustration. This could have been the reason behind multiple registrations by the same person. Furthermore, we should not assume that individuals have the same familiarity with novel interaction mechanisms and objects, as they do, say, with keyboards and characters [46].

VII. REFLECTION

We detailed Tetrad's deficiencies in the previous section as well as the difficulties we faced because of the nature of the evaluation. The question is: "were these revealed because of the field test, or could they have been detected by a lab study?"

Setting up the field test was a huge task that we conducted in good faith, believing it to be superior for testing Tetrad's smartphone implementation rather than a further lab study. In terms of cost-benefit, was it worthwhile?

Consider each of the deficiencies that emerged during the field trial: Effectiveness, Efficiency, Satisfaction and Interface Inconsistency. The predictability of image choice is a well-known problem for graphical authentication mechanisms. We could undeniably have trialled our image choice in a lab study. The inconsistency of the interface, too, would have been revealed during a simple usability evaluation. The inconvenience of authentication could also easily have been measured. The usefulness of the application has nothing to do with the authentication mechanism and this deficiency is a direct consequence of the decision to carry out a field test while wanting to ensure that evaluators were not put at risk. This meant we couldn't feasibly protect anything of real value with Tetrad, and this constraint meant the application was probably not worth the effort of authenticating.

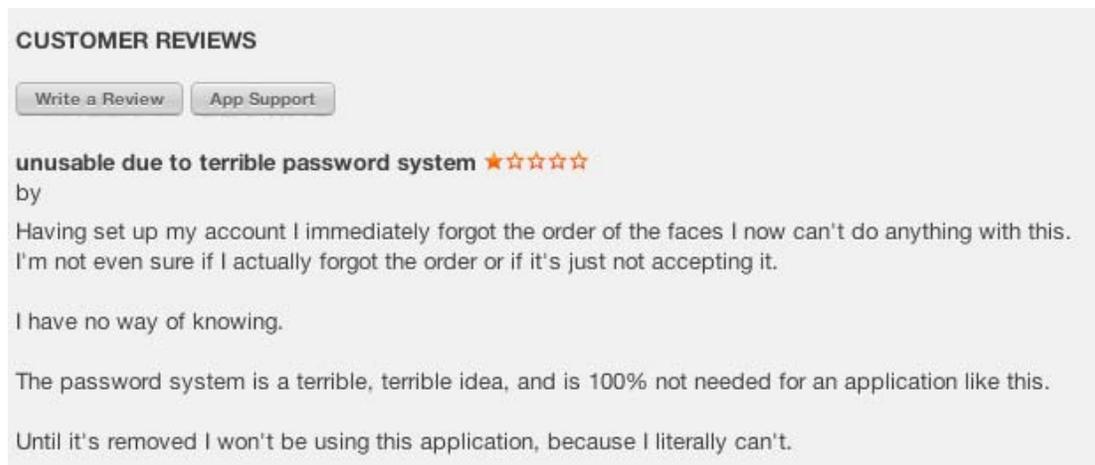

Fig. 3. iTunes Store review submitted by a participant.

So, in effect, we gave our participants an Application that did not meet their needs, and made them use a very inconvenient mechanism to access it. Moreover, the authentication mechanism did not offer a high degree of security since the image choice was predictable. This did not constitute a particular issue since what was being protected (lecture recordings) were freely available from the school's website. Here was a further issue: we put a hurdle in the way of their accessing something they could access very easily using another channel. We can only conclude that we ought not to have carried out this field test.

If we ignore the flawed nature of the application for the moment, let us focus on the purpose of our field test: evaluating the mechanism itself. It could be that our Tetrad smartphone implementation was too poor to justify probationing it. The initial lab evaluation did not reveal the problems we experienced during the lab study, probably due to the fact that in the lab study authentication was the main task and not an obstacle to be hurdled to access an application. Moreover, the initial application was implemented for a large screen and it was perhaps the novelty of the mechanism that attributed to the positive responses. It could have been that the mechanism was simply unsuitable for use on a smartphone.

We initially attributed the failure during the field test to these facts. We then started to wonder whether field testing is indeed the silver bullet it is accepted to be within this field. It is only worth carrying out if it delivers insights commensurate with the huge expense that goes into launching and sustaining it. Since we could have gained insights into the deficiencies of our mechanism in lab studies it clearly did not justify the expense in this case.

Are field tests still wise? They do seem to be the gold standard of evaluations. However, we strongly urge that multiple lab evaluations be carried out before a field trial is contemplated so that others do not go to huge expense, as we did, and have very little to show for it once the test has concluded. What we concluded at the end of this process was that we rushed into the field test prematurely.

## VIII. CONCLUSION

It would have been nice to have been able to report that Tetrad fulfilled its original promise during probation. It would have been a most satisfactory conclusion. Research, however, often delivers unexpected results, and this has certainly been the case here. Even so, we are reporting our results so that other researchers will be warned about auditioning their own mechanisms prematurely: so that others may avoid the errors we made.